\documentstyle{article}

\newcommand {\apj}{Astrophys. J.}
\newcommand{\aap}{Astron. Astrophys.}
\newcommand{\mnras}{Mon. Not. R. Astron. Soc.}
\newcommand{\prl}{Phys. Rev. Lett.}
\newcommand{\prd}{Phys. Rev. D}

\newcommand{\npa}{Nucl. Phys. A}

\usepackage{psfig, epsfig,rotate,multicol}
\begin{document}

\title{An Astronomical Evidence of Existence of Quark Matter   and  the Prediction for Submillisecond Pulsars}
\author{Zheng Xiaoping, Pan Nana, Yang Shuhua, Liu Xuewen, Kang Miao
\\
{\small 1 Department of Physics, Huazhong Normal University}\\
\small 2 Institute of Astrophysics, Huazhong Normal
University}
\date{}
\maketitle
\begin{abstract}
We derive the bulk viscous time scale of neutron stars with quark matter core, i.e. hybrid stars. The r-mode instability
windows of the stars show the theoretical result accords with the rapid rotation pulsar data. The fit gives a strong
indication for the existence of quark matter  in the interior of neutron stars. Hybrid stars instead of neutron or
strange stars may  result in submillisecond pulsars if they exist.

\end{abstract}
Searching for quark-gluon plasma(QGP) and quark matter stars are of great interest in high energy physics
as well as astrophysics. Compact objects are thought to  provide a unique astrophysical environment for
 the transition from  hadron matter into quark matter. A neutron star with quark matter core is usually called
 hybrid star.  However the prediction of distinguishable signature from neutron stars is urgent and difficult task
because of similar in mass and size for the two kinds of stars and their same surface properties as well.
An alternative way is to clarify limiting
rotation constrained
from the  dynamics of dense matter in the interior of the stars. Recently the discovery of r-mode instability
in relativistic stars\cite{and98} gives us  a chance to realize this issue.
 A series of
papers have investigated the many implications for gravitational
radiation detection and the evolution of pulsar\cite{fri98,
lin98, owe98, koj98, mad98, and99a, and99b, ho00, rez00,
mad00,and02}. Two  crucial issues regarding
the star evolutionary relevance of the r-mode instability also have been
solved: key results concern the interaction between
oscillations in core fluid and the crust\cite{bil00, and00,
lin00a} and  bulk viscosity coupled to r-modes\cite{lin99}.

We have known two evident characteristics regarding rotation of pulsars from the observed data, the most rapidly spinning
pulsar with 1.5ms period and the clustering of spinning frequencies of the low mass x-ray accretion binaries(LMXBs).
 Many investigators have made efforts to recognize the observable properties or what they indicate in neutron star
 or strange star models through the studies on so called r-mode instability window. In the beginnings,  some works find
  hot, young neutron stars would be  spun down to 15ms-20ms from their birth due to gravitational wave radiation because of
  r-mode unstable\cite{and99a}. Soon after,  neutron stars with solid crust could be thought a candidate model to explain
  the rapidly rotating accreting pulsar's data   through thermal runaway recycle of r-mode
  when the viscous boundary layer damping  taken account into\cite{bil00, and00}. Meanwhile, it was proposed that newly born
strange stars would also be spun down to about 2.5ms-3ms due to a huge bulk viscosity relative to neutron star
matter\cite{mad00}.

However, the resulting viscous heating in the boundary layer for neutron stars with solid crust
is so intense that it can heat the crust-core interface to
the melting temperature of the solid crust if the r-mode amplitude is larger than some critical value, which is crudely
estimated to be $\sim 10^{-3}$ by Owen\cite{owe} and  a more accurate value is given as $5\times 10^{-3}$
 by Lindblom et al.\cite{lin}. Therefore Lindblom et al. found it appear unlikely in neutron stars
 that the r-mode instability is responsible for limiting the spin periods of LMXBs\cite{lin}.
For strange stars, the spin-down to about 3ms corresponds to temperature in $10^6--10^7$K instead of  the inferred  core temperature
as a few times $10^8$K, which is a serious drawback. It isn't genuinely convincing  if the drawback  is simply
 attributed to under-representing statistics due to a low number of objects\cite{mad00}.

 In this paper, we will focus on the r-mode instability window of neutron stars with quark matter core and investigate
 the spin-evolution of the stars in presence of r-mode instability . As demonstrated below,
 the rapidly rotating pulsars would be a strong
 indication of the existence of quark matter in the interior of neutron stars(i.e.,  hybrid stars).
 A hybrid star would also be  inferred to  a submillisecond pulsar  if the quark matter core is large enough.

In rotating relativistic stars, gravitational radiation drives the r-mode while various dissipation mechanisms counteract
the fluid motion. In general, shear viscosity, or surface rubbing if the star has a solid crust, suppresses the mode
at low temperatures but bulk viscosity dominates at high temperatures. The critical rotation frequency for a given stellar model as a function of temperature follows from
\begin{equation}
{1\over\tau_{gr}}+{1\over\tau_{sv}}+{1\over\tau_{bv}}+{1\over\tau_{sr}}=0,
\end{equation}
where $\tau_{gr}<0$ is characteristic time scale for energy loss due to gravity wave emission, $\tau_{sv}$ and $\tau_{bv}$
are the damping times due to shear and bulk viscosities, $\tau_{sr}$ is surface rubbing due to the presence of the viscous
boundary layer.

A polytropic equation of state with a low index $n$ is a good approximation for compact stars, as was discussed in many
papers\cite{kok99,lin99}. The  time scale for $n=1$ polytrope from gravity wave emission can be written
\begin{equation}
\tau_{\rm gr}=-3.26({\rm\pi}G\bar\rho/\Omega^2)^3
\end{equation}
where $G$ is the gravitational constant, $\Omega$ is angular rotation frequency, and $\bar\rho$ is the mean density. Since
both the shear viscous damping  and the  rubbing dissipation due to viscous boundary layer dominate at low temperatures
and some investigations have shown  $\tau_{sv}\gg\tau_{sr}$\cite{bil00,and00}, only the surface rubbing need to be considered. In
ref\cite{and00}, the time scale is estimated as
\begin{equation}
\tau_{sr}=200\times\left(M\over 1.4M_\odot\right)\left(R\over 10{\rm km}\right)^{-2}\left(T\over 10^8{\rm K}\right)\left(P\over 1{\rm ms}\right)^{1\over 2}
\end{equation}

We now need to evaluate the bulk viscous damping time scale of hybrid stars.  We can follow \cite{lin99} to work out
 bulk viscosity coupled to r-modes taking respectively the viscosity coefficients of strange quark matter(SQM) and neutron star matter
  in the core and the outer covering of the star, but the calculation is a complicated and arduous task.
  In fact, we will only consider the dissipation on r-modes in quark matter core because of a huge bulk viscosity
 of SQM relative to neutron star matter. As demonstrated below, we  can here avoid  the complicated calculations to
  deduce the time scale with a simple method following the formula in \cite{lin99}.

  For the given energy of modes $E$ and its variable rate $ \dot{E}\equiv {\rm d}E/{\rm d}t$, the time scale is defined as
  \begin{equation}
  {1\over\tau}=-{\dot{E}\over 2E}
  \end{equation}
  The r-mode energy in a rotation relativistic star with radius $R$ has been expressed as
  \begin{equation}
  E={\alpha^2\pi\over 2m}(m+1)^3(2m+1)!R^4\Omega^2\int_0^R\rho(r)\left ({r\over R}\right )^{2m+2}{\rm d}r+O(\Omega^4),
  \end{equation}
  where $\alpha$ is the dimensionless amplitude,  $m$ is natural number and $\rho(r)$ is the density function.
  The energy dissipation rate due to bulk viscosity can be evaluated with the
  following formula
  \begin{equation}
  \dot{E}_{bv}=-\int\zeta\delta\sigma\delta\sigma^*{\rm d}^3x,
  \end{equation}
where $\zeta$ is bulk viscosity coefficient, $\delta\sigma$ is deformed velocity tensor.
We first decompose  the energy $E$ into two parts, i.e., the quark core and nuclei envelope denoted respectively by
$c$ and $e$
\begin{equation}
E=\left ({R_c\over R}\right )^{2m-2}(E_c+E_e)
\end{equation}
with
\begin{equation}
E_c={\alpha^2\pi\over 2m}(m+1)^3(2m+1)!R_c^4\Omega^2\int_0^{R_c}\rho_c(r)\left ({r\over R_c}\right )^{2m+2}{\rm d}r,
\end{equation}
\begin{equation}
E_e={\alpha^2\pi\over 2m}(m+1)^3(2m+1)!R_c^4\Omega^2\int_{R_c}^R\rho_e(r)\left ({r\over R_c}\right )^{2m+2}{\rm d}r.
\end{equation}
Assuming uniform core and envelope that is approximately correct for neutron stars, we immediately estimate
\begin{equation}
{E_e\over E_c}\approx {\rho_e\over\rho_c}\left ({R\over R_c}\right )^{2m+3}.
\end{equation}
Thus the energy can reads
\begin{equation}
E=\left ({R_c\over R}\right )^{2m-2}\left [ 1+{\rho_e\over\rho_c}\left ({R\over R_c}\right )^{2m+3}\right ]E_c.
\end{equation}
An analogous treatment can be made for the energy dissipation, but we take the estimation,
${\rm d}\sigma\sim \left ({R_c\over R}\right )^{m-1}R_c^2\left ({R_c\over R}\right )^{m+1}$, into consideration according \cite{lin99}.
Regarding the fact that the SQM bulk viscosity dominates\cite{mad92}, i.e., $\zeta_{\rm SQM}\gg \zeta_{\rm NSM}$, we easily obtain the
dissipation due to bulk viscosity
\begin{equation}
\dot{E}_{bv}\approx\left ({R_c\over R}\right )^{2m-2}\dot{E}_{Bc}.
\end{equation}
Therefore we derive the time scale as
\begin{equation}
\tau_{bv}=\left [1+{M-M_c\over M_c}\left ({R\over R_c}\right
)^{2m}\right ]\tau_{Bc},
\end{equation}
where $M_c$ represents the mass of SQM core in the interior of
stars. Evidently, the time scale $\tau_{Bc}$ is in agreement with
that of a strange star with radius $R_c$, which can be
calculated\cite{lin99}. For given $R, R_c, M, M_c$(10km, 5km,
1.4$M_\odot$, 0.7$M_\odot$, we obtain
\begin{equation}
\tau_{bv}=5.4\times 10^2{\rm s}\left({M\over 1.4 M_\odot}\right )\left({R\over 10{\rm km}}\right )^{-4}
\left({T\over 10^9{\rm K}}\right )^{-2}\left ({m_s\over 100{\rm MeV}}\right )^{-4}\left ({1000\over\nu}\right )^2
\end{equation}
where $m_s$ is strange quark mass. In fact, the reasonable radius
of quark matter core in interior of  hybrid stars is able to range
from 3km to 6km\cite{sch}. The time scale strongly depend on the
radius of the quark matter core from Eq(13),it is in the range of
 $2.4\times 10^3$s - $4.7\times 10^1$s
under the situation of the  scaling quantities given in the above formula.


To have those time scale quantities, we can apply equation(1) to
evaluate the critical  frequency of the rotating stars as a
function of temperature. Figure 1 plots  the instability window
for the given parameters: $R=10$km,$R_c=5$km, $M=1.4M_\odot,
M_c=0.7M_\odot, m_s=200$MeV. The result(solid line) in figure 1 shows that the
remain pulsars in supernova explosions will reach the instability
window due to cooling and then slow down by gravitational wave
emission braking (high braking index 9) along the right edge of
the window. This is similar to  bar strange stars but the stars
here  only spin down to 704Hz (a period of 1.42ms). Especially,
the lowest position of the window can be located at  the inferred
star core temperature, truly consistent with the data unlike
situation of bar strange stars. In contrast to neutron
star(without quark matter core), our new results show the pulsars
follow a track indistinguishable from the critical rotating curve
govern by the competition between bulk viscous dissipation and
gravitational wave emission to be spun down according to the
mechanism discussed by Andersson et al.\cite{and02} and
Madsen\cite{mad00}. Obviously, the solid crust in hybrid stars
will never melt even if the stars accreting spin-up. To display
clearly hybrid star advantage, we have also collected the results
for neutron stars and bar strange stars in figure 1. Finally, it
is worthwhile discussing the dependence of limiting rotation of
pulsars on the size of quark matter core. Figure 2 shows the
limiting rotation frequency of pulsars as hybrid stars in the
range of 637-826Hz in temperature region of $1\times 10^8-4\times
10^8$K,
 based on supposition that the radius of the quark matter core  takes 3-6km, as seen above, from nuclear theoretical prediction\cite{sch}.
The distributed data region are full in agreement with our those results.


In addition, we can imagine that the existence of the higher rotation pulsars is possible if there would exist the neutron stars
with larger quark matter core. The thick line in figure 1 indicates the upper limit of the rotation of the pulsars as hybrid stars,
which is higher than neutron stars and strange stars and over kHz (a period of submillisecond). This implies that
submillisecond pulsars, if they exist, should be hybrid stars instead of neutron stars or  strange stars, which needs future
study.

 In conclusion, the bulk viscous time scale for hybrid stars has been calculated for a range of parameters. The r-mode
 instability window has been obtained. The rapid rotation pulsar data are a significant indication of the existence of
 hybrid stars instead of neutron stars or  strange stars. Of course, this is a possible evidence of the existence of Quark Matter
  in the interior of neutron stars. Meanwhile, hybrid star model may predict the existence of submillisecond.

  We would like to thank the  support by National Natural
Science Foundation of China under Grant No.


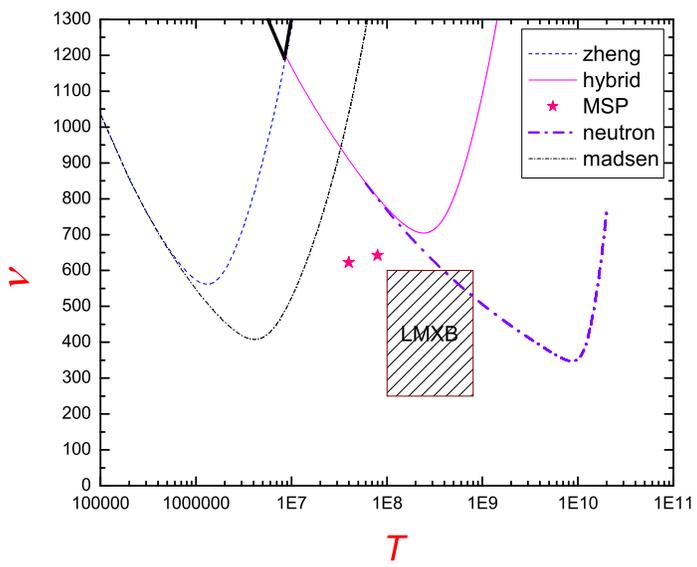
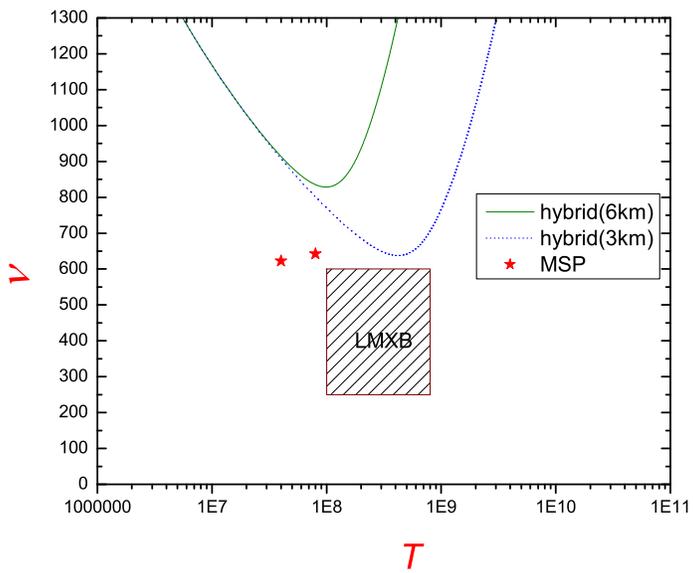
\end{document}